\documentclass[letterpaper,12pt]{article}

\usepackage{fullpage}
\usepackage{amssymb}
\usepackage{xspace}
\usepackage{hyperref}
\usepackage{graphicx}

\newcommand{\TeV}{\ensuremath{~\mathrm{TeV}}}
\newcommand{\HT}{\ensuremath{H_\mathrm{T}}}
\newcommand{\pT}{\ensuremath{p_\mathrm{T}}}
\newcommand{\MD}{\ensuremath{M_D}}
\newcommand{\Ms}{\ensuremath{M_\mathrm{s}}}
\newcommand{\gs}{\ensuremath{g_\mathrm{s}}}
\newcommand{\Mth}{\ensuremath{M_\mathrm{th}}}
\newcommand{\sumpT}{\ensuremath{\sum p_\mathrm{T}}}
\newcommand{\charybdis}{{\tt CHARYBDIS2}\xspace}
\newcommand{\blackmax}{{\tt BlackMax}\xspace}
\newcommand{\qbh}{{\tt QBH}\xspace}

\begin{document}

\begin{center}
\LARGE\bfseries
Collider searches for non-perturbative low-scale gravity states
\end{center}  

\bigskip

\begin{center}
Douglas M. Gingrich\\

\bigskip

\textit{Centre for Particle Physics, Department of Physics, University
of Alberta, \\ Edmonton, AB T6G 2E1 Canada}\\  
\textit{TRIUMF, Vancouver, BC V6T 2A3 Canada}\\ 
{\footnotesize gingrich@ualberta.ca}
\end{center}

\begin{center}
\small \today
\end{center}

\begin{quotation} \noindent
\textbf{Abstract\ } 
The possibility of producing non-perturbative low-scale gravity states
in collider experiments was first discussed in about 1998.  
The ATLAS and CMS experiments have searched for non-perturbative
low-scale gravity states using the Large Hadron Collider (LHC) with a
proton--proton centre of mass energy of 8\TeV. 
These experiments have now seriously confronted the possibility of
producing non-perturbative low-scale gravity states which were
proposed over 17 years ago.  
I will summarise the results of the searches, give a personal view of
what they mean, and make some predictions for 13\TeV{} centre of mass
energy. 
I will also discuss early ATLAS 13\TeV{} centre of mass energy results.
\end{quotation}

\begin{quotation} \noindent
\end{quotation}

\newpage
\tableofcontents
\newpage

\section{Introduction}

Brane world
scenarios~\cite{ArkaniHamed:1998rs,Antoniadis:1998ig,Randall:1999ee} 
offer paradigms to reinterpret the four-dimensional Planck scale
$M_\mathrm{P}$ as an effective gravity scale arising from a more
fundamental lower gravity scale $M_*$ in higher dimensions. 
This reinterpretation of the gravity scale allows new phenomenological 
models~\cite{Giddings:2001bu,Dimopoulos:2001hw} to be developed which
help guide searches for low-scale gravity in experiments, such as
those at the Large Hadron Collider (LHC).    
An exciting outcome of these models is the possibility to produce
non-perturbative gravity states at the LHC.
The ATLAS and CMS experiments have recently published a round of
searches at 8\TeV{} proton--proton centre of mass energy for
non-perturbative gravity states which seriously confront the models for
the first time. 
I examine how the models can now be viewed in light of the experimental 
constraints.

\section{Non-perturbative gravity states}

Before we begin, I discuss how including non-perturbative gravity into
particle physics can perhaps involve a slight modification to the usual 
thinking. 
The way of thinking is slightly different from main-stream particle
physics. 
Particle physicists are use to searching for new particles.
They need quantum mechanics and special relativity to described them.
For calculations, they usually have a Lagrangian in field theory, and
use perturbative techniques to expand the result in a series of
Feynman diagrams.  
States with energy above the gravity scale (transplanckian scale physics)
are described non-perturbatively.
Classical or semi-classical mechanics should hold.
Being non-perturbative, expansions in a coupling constant and Feynman
diagrams do not make much sense.

Like searches for new particles, we usually think of one force -- in this
case gravity -- dominating the interaction and ignore the others -- in
this case QCD -- so many QCD issues (LO, NLO, NNLO, etc.) are not
relevant for non-perturbative gravity states. 

Several paradigms for model development exist.
I refer to them as paradigms as they are not specific models but
frameworks which allow models to be developed.
Two extra dimensional scenarios are the most popular:
large flat extra dimensions proposed by Arkani-Hammed, Dimopoulos,
Dvali (ADD)~\cite{ArkaniHamed:1998rs,Antoniadis:1998ig}, and a warped
extra dimension in AdS space proposed by Randall and Sundrum
(RS1)~\cite{Randall:1999ee}. 
Universal extra dimensions are also popular but not relevant to the
phenomenology discussed here.
An alternative approach to using extra dimensions to lower the Planck
sale is due to Dvali, in which a large number of particle species 
(messenger particles) are proposed~\cite{Dvali:2007wp}.
In general, one needs some idea to reduce the Planck scale $M_\mathrm{P}$
to a lower gravity scale $M_*$, such that $M_\mathrm{P} \gg M_*$.
I will use \MD{} for $M_*$ from now on, although the definition of the
scale is model dependent. 

Many ideas describing the effects of low-scale gravity exist but few of
them allow a concrete model to be developed and utilised by experiments
to perform a search for low-scale gravity phenomena.
The most popular model describes higher-dimensional black holes using
the theory of general relativity.
This model treats the production of black holes in particle collisions
classical, and the decay is teated using the semi-classical physics of
Hawking evaporation~\cite{Hawking75}.
Throughout this review, I will refer to these types of black holes
models as general relativistic (GR) black holes.

Since string theory, such as superstrings, occurs in higher dimensions,
it is natural to attempt to embed string theory into ADD.
In the regime of weakly-couple string theory, the correspondence between
black holes and string states manifests itself in terms of a
highly-exited string state, or string ball~\cite{Dimopoulos:2001qe}.
I will refer to string balls together with GR black holes, as thermal 
black holes.

The above two models consider the non-perturbative objects as thermal
states. 
It is believed in the quantum gravity regime the transplanckian gravity
state will behave more like a particle, and decay
non-thermally~\cite{Meade08}. 
These non-thermal black holes are often called quantum black holes or
QBH. 
I will refer to them as non-thermal black holes. 

A few other less popular models have been discussed in the context of
having different phenomenology to the above at the LHC.
Attempts to calculate the gravity cross section of two colliding
particles (trapped-surface calculations) allows alternative levels for
black hole production~\cite{Eardley02}.
The early concept of fermions residing on separate
branes~\cite{Arkani99p}, split-fermions,  makes concrete predictions and
can be used in model building. 
ADD has been embedded into a non-commutative geometry~\cite{Rizzo06b} in
hopes to better model the effects of quantum gravity.

A model is of little use to an experiment unless it can be implemented
in a Monte Carlo event generator. 
One such generator is \charybdis~\cite{Frost:2009cf} which simulates GR
black holes. 
String balls have been added~\cite{Gingrich:2008di}, and the code
modified for non-commutative black holes~\cite{Gingrich:2010ed}.
The \blackmax{} generator~\cite{Dai:2007ki} also simulates GR black holes,
has had string balls added~\cite{Gingrich:2008di}, and is capable of
simulating split-fermion models. 
The \qbh generator is most often used for simulating non-thermal black
holes~\cite{Gingrich:2009da}.

To discuss searches for non-perturbative gravity states, we need
to first access the current bounds on the parameters of the models.
I will restrict this discussion to ADD.
No bounds exist on the number of extra dimensions besides that one large
flat extra dimension is not consistent with our daily observations.
The parameter of interest is the fundamental Planck scale \MD{} and I
will ask what the limits on this parameter are from existing
experimental measurements.
An up to date summary can be found in Ref.~\cite{Gingrich:2012vs}.
Searches for virtual graviton emission depend on an ultra-violet cutoff
\Ms, which is not \MD.
Real graviton emission depends on \MD.
The most stringent limits come from searches for mono-jet events and
mono-photon events.
But is this the scale for thermal and non-thermal black holes?
I argue that this is the case when the search is interpreted in terms 
of the ADD model, i.e. the same model that is predicting the
non-perturbative gravity states.
Direct searches for non-perturbative gravity states do not allow
very stringent limits on the fundamental Planck scale.
Limits on thermal states are given in terms of \MD{} as function of
\Mth, a mass threshold.
Since \Mth{} is not a physical parameter, it is not possible to infer
much about \MD{} from these searches.
Limits from non-thermal black hole searches can assume $\MD =
\Mth$, where a limit is set on the later quantity.
Since this assumption is only approximately valid, limits on
\Mth{} are not valid statistical limits on \MD.
In the following discussions, I will use the most stringent limits of
$(n,\MD) = (2,5.61\TeV)$,  $(3,4.38\TeV)$, $(4,3.86\TeV)$,
$(5,3.55\TeV)$, and $(6,3.26\TeV)$ from the CMS mono-jet
search results~\cite{Khachatryan:2014rra}.

\section{Searches for non-perturbative gravity states}

Only the LHC experiments ATLAS and CMS have performed searches for
non-purtubative gravity states.
In the discussion of these results, I will divide the searches into
thermal and non-thermal gravity states.
For each experiment, I consider only the most recently published paper
using a particular analysis strategy.
These sometimes supersede previous publications which used data at
lower energy or lower luminosity, or results found in LHC public
conference notes. 

Thermal gravity states such as GR black holes and string balls
have been searched for by ATLAS~\cite{Aad:2015zra} and
CMS~\cite{Khachatryan:2010wx,Chatrchyan:2012taa,Chatrchyan:2013xva} in
multi-jet events.  
In addition, ATLAS has search for the same states in $\ell+$jets
events~\cite{Aad:2012ic,Aad:2014gka}, and same-sign dimuon events with
a large number of tracks~\cite{Aad:2011bw,Aad:2013lna}. 

Non-thermal black holes have been search for by
ATLAS~\cite{Aad:2014aqa} and
CMS~\cite{CMS:2012yf,Chatrchyan:2012taa,Chatrchyan:2013xva,Khachatryan:2015sja}
in dijet events. 
In addition, ATLAS has searched in $\gamma + $jets
events~\cite{Aad:2013cva}, $\ell+$jets events~\cite{Aad:2013gma}, and
dilepton events~\cite{Aad:2014cka} for non-thermal black holes.
In all cases, I refer to a lepton as either an electron or a muon only.
ATLAS has also searched in dijet events for thermal black holes
extrapolated down to the Planck scale~\cite{Aad:2011aj,ATLAS:2012pu,Aad:2014aqa}.

\subsection{Searches for GR black holes}

The key feature of GR black holes is that they are thermal states
which Hawking evaporate.
The evaporation is a semi-classical description and the production
mechanism is described classically.
However, since low-scale gravity is expected to require a quantum
mechanical description, we need to introduce a cut-off parameter
$M_\mathrm{th}$ and impose the condition $E > \Mth \gg \MD$ for the
GR black hole models to be valid.  
Thus these states offer no predictive power of what we would see first
at lower energies ($E < \Mth$) at the LHC if low-scale gravity is
realised.  
For first signs of low-scale gravity, it is best to look for
perturbative states, such as Kaluza-Klein resonances, graviton
scattering, etc. 

Thermal decays are anticipated to give rise to mostly partons, which
will then hadronise and create jets.
Thus a search in events with a high multiplicity of high-\pT{} jets (or
particles) is a good choice for detecting thermal black holes.
In this type of search, the QCD background can be high, but can be
reduced by requiring a high-\pT{} lepton in the event.
A significant fraction of leptons should occur in high-multiplicity
thermal decays.
In these searches a non-physical mass threshold \Mth{} is introduced to 
keep the black hole classical. 
As we shall see, the criteria of $\Mth \gg \MD$ is seldom obeyed in the
stated experimental limits, and hence the results are of limited use in
constraining these model. 
This was also pointed out by Park~\cite{Park:2011je}.


Model-dependent limits have been set in multi-jet searches by
ATLAS~\cite{Aad:2015zra} and
CMS~\cite{Khachatryan:2010wx,Chatrchyan:2012taa,Chatrchyan:2013xva},
and in $\ell+$jets~\cite{Aad:2012ic,Aad:2014gka} and
dimuon~\cite{Aad:2011bw,Aad:2013lna} searches in ATLAS. 
The models used are given by rather standard configurations of the
\charybdis{} and \blackmax generators.
A two-dimensional parameter space in \MD{} and \Mth{} is used, thus
giving 95\% confidence level (CL) contour limits on these parameters.
I remind the reader that \Mth{} is not a physical parameter of the
model. 
The region in the bottom left of the plots (low \MD{} and low \Mth) is
excluded. 
These limits are obtained by assuming the model production cross
sections to have no uncertainty. 
Sometimes uncertainties in the parton density functions are taken into
account in the efficiency calculation. 
Since black hole cross sections are highly speculative these limits can 
only represent limits on some assumed model for the cross section, and
should not be considered limits on new fundamental physics.

Limits on \Mth{} are set for a series of \MD{} values.
The \MD{} values range from 1.5\TeV{} to 4.5\TeV{} in CMS and 1.5\TeV{} to
4.0\TeV{} in ATLAS. 
The lower bound of 1.5\TeV{} seems rather low when considering the limits
on \MD{} from searches for graviton emission, and all the other exotics
and SUSY searches performed at the LHC, which have not seen any hint for
new physics, let alone, a new physics scale.
The upper bound on \MD{} seems rather arbitrary. 
Perhaps the reason to not go higher has been to avoid a region in which
the models are not valid, but I will argue below that the models are
unlikely to be valid in any region of these contours. 
Park~\cite{Park:2011je} has expressed similar concerns four years ago.

\begin{figure}[htb]
\setlength{\unitlength}{0.07cm}
\begin{center}
\begin{picture}(90,170)
\put(10,10){\vector(1,0){50}}
\put(10,10){\vector(0,1){50}}
\put(62,9){$\MD$}
\put(8,62){$\Mth$}
\put(30,10){\line(0,1){150}}
\put(10,10){\line(1,5){30}}
\put(25,4){$\MD$(limit)}
\put(8,80){$k=5$}
\put(48,144){allowed region}
\put(46,145){\vector(-1,0){12}}
\put(35,110){$\Mth > 16\TeV$}
\end{picture}
\end{center}
\caption{GR black hole search parameter space.
The valid region is to the right of the vertical \MD(limit) line and to
the left of the $k=5$ line.}
\label{fig1}
\end{figure}
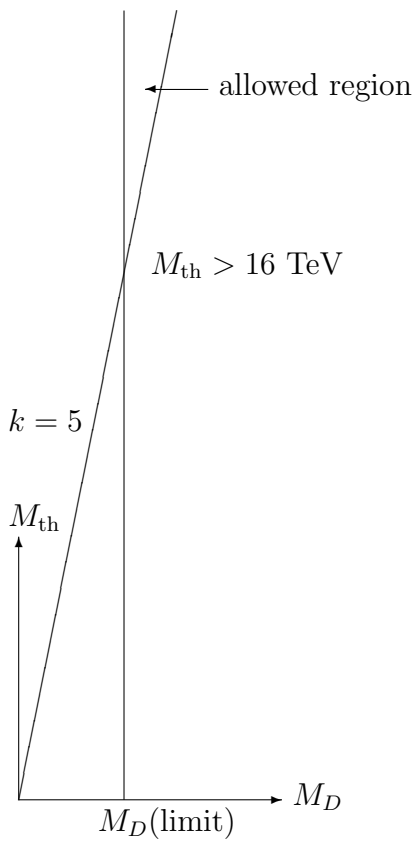

Figure~\ref{fig1} shows a schematic of the model-dependent limit
validity region.
For a given value of $n$, limit contours are set in $(\Mth,\MD)$.
Based on existing lower limits on $\MD$, the region to the left of the
$\MD$(limit) vertical line is excluded.
A metric for the validity of classical models is often given as $\Mth
\gg \MD$, or $k \equiv \Mth/\MD \gg 1$.
Clearly this is not the case in the published contour limits.
A popular value for validity in the literature is $k>5$, which is also
questionable.
This results in a region of validity to the left of the $k=5$ line in
Fig.~\ref{fig1}. 
The allowed region is thus $\Mth > 16$\TeV, which is currently beyond
the LHC energy reach. 
In summary, the model-dependent limits to date are not particularly 
relevant to constraining low-scale gravity physics.

The limits can provide some comparisons between models, analysis
procedures, and differences between experiments.
The limits on \Mth{} are typically highest at low \MD{} and decrease
monotonically with increasing \MD.
Comparing the multi-jet analyses to the $\ell+$jets analysis you
can see that the lower efficiency for $\ell+$jets is dominating over  
the reduction in QCD background to give slightly less stringent limits. 
You can also see that the $\ell+$jets limits are more dependent on  
rotation and decrease faster as \MD{} increases.
This is related to the requirement of a high-\pT{} fermion in the final
state. 
The multi-jet analysis of both ATLAS and CMS are comparable.
The small differences can be attributed to the small luminosity
difference and the inclusion of missing transverse momentum by CMS.

I use the following procedure when discussing the mass limits.
The cross section limits are interpret with models that are generally a
function of $(n,\Mth,\MD)$.
$n$ is a fundamental parameter of the brane-world and can only have one
value. 
Limits have been set for $n=1-6$ (CMS), but ATLAS considers only $n=2$
and 6, at most. 
We have no information for larger $n$, and possibilities are not ruled
out in non-string theory models.
Limits on \MD{} for different $n$ have been
obtained~\cite{Khachatryan:2014rra}.
I invoke these limits as a condition to reduce the $(\Mth,\MD)$ limit
space.  
This gives limits on \Mth{} for different $n$.
I also note the $k$ value and suggest that it be reasonable for the  
model used.

All searches for GR black holes have set model-dependent limits.
However, recent results from ATLAS using 80~pb$^{-1}$ of data with the
LHC running at 13\TeV{} centre of mass energy have set the the most
stringent limits.
A multi-jet analysis~\cite{atlas:run2multijet} obtains $\Mth >
8.5-7.5$\TeV{} for $\MD = 2-5$\TeV{} ($k = 4.2-1.5$) at the 95\% CL. 
Invoking the current limits on \MD{} gives $\Mth > 8.1$\TeV{} ($k =
2.5$). 
Similarly, a $\ell+$jets analysis~\cite{atlas:run2lplusjet} obtains
$\Mth > 7.3-5.9$\TeV{} for $\MD = 2-4$\TeV{} ($k = 3.6-1.5$) at the 95\% 
CL.
Invoking the current limits on \MD{} gives $\Mth > 6.4$\TeV{} ($k = 2$).
While these are significant improvements over the mass threshold
limits at 8\TeV{} proton--proton centre of mass energy, they are still
not in a region of parameter space in which the models are
particularly valid. 

\subsection{Searches for string balls}

Embedding weakly-coupled string theory into ADD results in string ball 
states that could be searched for at the LHC~\cite{Dimopoulos:2001qe}.
The model~\cite{Gingrich:2008di} modifies the black hole cross section,
but leaves the decay mechanism similar to thermal black holes --
except the temperature is different.  
String ball models are expected to be valid at lower values of energy
then the GR models.
This is accomplished by introducing another scale (the string scale \Ms)
that allows $E > \Mth \gg \Ms$ and $\MD > \Ms$.
In reality, this just pushes the validity of GR black holes to lower
energy at the expense of more speculation (low-scale string theory).   


Similar to GR black holes, model-dependent limits have been set for
string balls. 
String balls have the additional parameters of the string scale and
string coupling.
However, when also considering the Planck scale, only two of the three
parameters are independent.
The experiments have chosen to take \Ms{} and \gs{} as the independent
parameters. 
\gs{} is fixed as part of the model, and \Ms{} replaces \MD{}.
\MD{} is not independent and is calculated.
The contours are thus in \Mth{} and \Ms{} space.
\Ms{} ranges from 0.8\TeV{} to 3.0\TeV{} in ATLAS.
Although there have not yet been any direct limits on \Ms{} it is hard to
imagine from the many searches that have been performed at the LHC
that they have not ruled out values as low as 0.8\TeV.
The upper search bound on \Ms{} seems arbitrary.
A requirement of $k = \Mth/\Ms \gg 1$ is also necessary for validity of
the model.
This ratio could perhaps be lower than for the case of GR black holes.
A common choice is $k > 3$, and this is satisfied in a region of the
search space.
However, this is a region in which \Ms{} is low, so the valid region
exists where string physics has probably been excluded, and the region
of allowed \Ms{} is in a invalid region of the model.
In this regards, these limit contours are also of limited use for
constraining low-scale gravity.
Perhaps at higher LHC energy the limits will constrain the model.
When setting limits on string-ball models it is important to choose
the parameters such that the stringy-regime of the cross section is in
the region where the limits are set, else one is effectively setting
limits on GR black holes. 

String balls have been searched for in the multi-jet and $\ell+$jets
final states.
Both ATLAS and CMS set limits using a model in which $\gs = 0.4$.
The most stringent limits on string balls come from the multi-jet
searches. 
Taking rotating string balls as an example, the ATLAS
limits~\cite{Aad:2015zra} on (\Mth, \Ms) range from about (6.35\TeV,
0.8\TeV) ($k = 7.9$) to about (4.9\TeV, 3.0\TeV) ($k = 1.6$) at the
95\% CL.  
For $k = 3$, the lower limit is (5.35\TeV, 1.8\TeV).
CMS obtain similar limits~\cite{Chatrchyan:2013xva}.
The limits from the ATLAS search in the $\ell+$jets final state are
less stringent~\cite{Aad:2014gka}.  

\subsection{Model-independent limits}

The lack of discovery in signal regions allow upper limits to be set on
the number of signal events $N_\mathrm{upper}$, which is independent of
model assumptions. 
These limits can then be used for any new physics model that has the
same signature as that used in the search.
These upper limits on events allow a determination on the upper limit of
cross section $\sigma_\mathrm{upper}^\mathrm{vis}$ times branching
fraction $B$ times efficiency $\varepsilon$ times acceptance $A$:

\begin{equation}
N_\mathrm{upper} = \sigma_\mathrm{upper}^\mathrm{vis} L A
\varepsilon\, , 
\end{equation}

\noindent
where $L$ is the integrated proton--proton luminosity, which is
typically about 20~fb$^{-1}$ in the analysis discussed here.
This is often referred to as the visible cross section.
Thermal black hole searches are usually made using inclusive final
states so the branching fraction is included in the efficiency. 

\begin{figure}[htb]
\setlength{\unitlength}{0.1cm}
\begin{center}
\begin{picture}(130,80)
\put(10,10){\vector(1,0){100}}
\put(10,10){\vector(0,1){60}}
\put(112,8){$\sum\pT$}
\put(48,15){$\sum\pT^\mathrm{min}$}
\put(15,19){$\sigma_{95}^\mathrm{min}$}
\put(8,73){$\sigma_{95}$}
\qbezier(20,60)(25,25)(50,20)
\put(50,20){\line(1,0){50}}
\put(50,8){\line(0,1){4}}
\put(8,20){\line(1,0){4}}
\end{picture}
\end{center}
\caption{Schematic of the 95\% confidence level cross-section upper
limit versus $\sum\pT$ (model-independent limit).}
\label{fig2}
\end{figure}
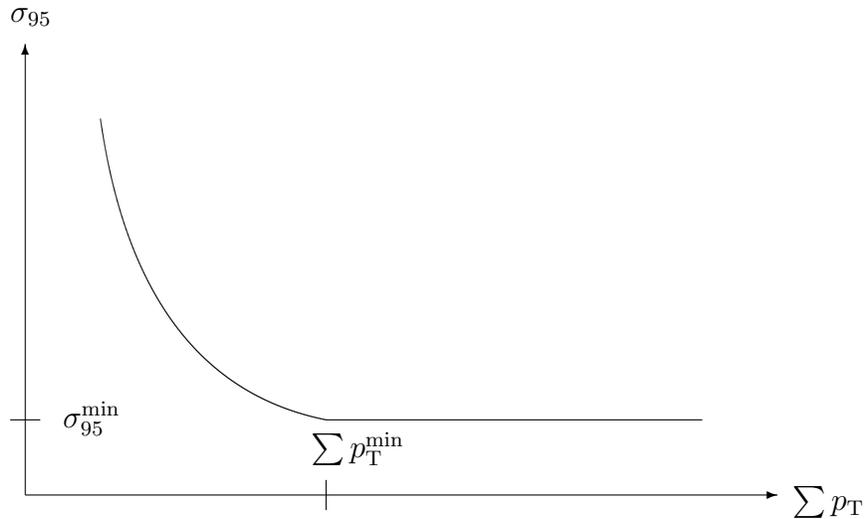

The cross section limits are presented as a function of a search
variable which is usually the scalar sum of the transverse momentum of
all the particles 
in the event \sumpT{}. 
For the multi-jet searches this is referred to as \HT.
This variable is often used inclusively, and the signal regions can
also be sliced in inclusive particle multiplicity.  
Figure~\ref{fig2} shows a schematic of the model-independent limit. 
Just knowing where in \sumpT{} the limit has a lower plateau is useful.
This represents the highest \sumpT{} value of all the events.
This then becomes a conservative and model-independent limit on
\sumpT{}, or whatever variable is used. 
Beyond this value of \sumpT{} there are no data events at higher
\sumpT{} and a negligible background in this region is predicted.
The plateau represents the highest excluded cross section, given by an 
upper limit of three events, above the highest $\sum\pT$ data event.

To set upper limits on the cross section requires the experiments to 
provide the acceptance time efficiency. 
To allow different models to be confronted, the acceptance is usually
not given, but is meant to be determined for the model of interest,
usually by Monte Carlo generator methods.
It is useful to make a clear distinction between the detector and
geometrical effects.
The efficiency on the other hand can only be determined by the
experiment.
In all but the simplest cases, this efficiency will depend on the model
considered. 
In this way the model-independent limits are not really model
independent. 

Sometime no efficiency is given, if the search is likely to be close to
fully efficient.
Sometimes the lowest efficiency is given, to be conservative, and will
probably allow estimates to a factor of a few.
Sometimes a mean and spread of efficiencies are given, or a set
of efficiencies.
One can then match the model of interest to the closest model for which
the efficiency is give.
Since models for non-perturbative gravity states are highly
speculative and uncertain, it should be sufficient to just take the
detector efficiency to by unity, and make an approximation of the
acceptance. 
In this way, the visible cross section is totally adequate for model
builders to determine if their model has already been rule out by
existing LHC experimental limits. 

The most stringent upper limits on the visible cross section are about
0.2~fb at the 95\% CL from the ATLAS multi-jet search~\cite{Aad:2015zra}.
This should be valid for most new phenomena resulting in high-\pT{}
and high multiplicity jet final states.
As a consequence, searches in run-2 at the LHC should not need to be 
concerned about signal contamination in the background estimate until
the data sample reaches a luminosity of about 15~fb$^{-1}$.

For classical black holes, the independent variable has been \sumpT.
Unfortunately, this variable is not related to analytical expressions for
the proton--proton or parton--parton cross sections.
Cross sections are usually given as a function of black hole mass $M$,
or the parton--parton centre of mass energy $\sqrt{\hat{s}}$.
There is no one-to-one relationship between \sumpT{} and $M$.
A transformation of \sumpT{} to $M$ will involve assumptions (model)
of the distribution of these variable, and unless the transformations
are done carefully, the result will not be a rigorous 95\% CL limit as
a function of $M$.  
However, since $\sumpT < M$ for well measured events, a limit in
\sumpT{} is a conservative limit in $M$.   
Exactly how conservative can only be approximated.
Removing the model-dependency and expressing the limit in $M$ would be a
great step forward.

\subsection{Searches for non-thermal black holes}

The LHC parton--parton centre of mass energy needs to be high relative
to \MD{} for a black hole to Hawking evaporate thermally -- so black
holes produced with a mass near \MD{} probably do not decay thermally
but decay more like a particle.
In the later case, considering a non-thermal black hole
model~\cite{Calmet08b,Gingrich:2009hj} is probably more
appropriate~\cite{Meade08}.  
Models for non-thermal black holes extrapolate the classical cross section
down to the Planck scale. 
More importantly, most models replace Hawking evaporation (thermal
decay) by particle decays. 
The branching fractions can be approximated by invoking conservation
principles.  
The ATLAS and CMS experiments have searched for non-thermal black holes in
a variety of two-body final states.

\subsubsection{Searches in the dijet final state}

Because gravity is expected to couple to all standard model particle
degrees of freedom equally, it is anticipated that the final states
involving quarks and gluons will dominate.
All 14 QBH states considered in Ref.~\cite{Calmet08b,Gingrich:2009hj}
will have a significant probability to decay to partons.
Such final states will lead to hadronic jets in the LHC detectors.  
Searching for new phenomenon in the dijet invariant mass spectrum is a
powerful approach to searching for non-thermal black holes.

Some of the first searches for new particles in the LHC experiments
interpreted the lack of a signal in terms of production limits on QBH
states. 
For CMS, the dijet signatures of QBH were initially included with the
thermal black hole searches~\cite{CMS:2012yf,Chatrchyan:2012taa} but
the latest results were
published~\cite{Chatrchyan:2013xva,Khachatryan:2015sja} along with
the search for narrow resonances in the dijet invariant mass spectrum.
CMS interpret the non-observation of a broad enhancement in the dijet
invariant mass spectrum as lower limits on the QBH mass of 5.0 to
6.3\TeV{} at the 95\% CL.
In obtaining this mass range, CMS consider $\Mth > \MD$ and take $\MD
= 2-5$\TeV{} and $n = 1-6$.
The limits range from about $(\Mth,\MD)$ of $(6.3\TeV,2.0\TeV)$
($k=3.2$) to about $(5.2\TeV,5.0\TeV)$ ($k=1.0$) for $n=6$ to $n=2$ at
the 95\% CL, respectively. 
Invoking the current limits on $\MD$ give limits of $\Mth >
5.6-6.05$\TeV{} ($k=1.4-1.9$) for $n=3-6$ at the 95\% CL.
This is the region in which the QBH model is expected to be valid.
The limits obtained for $n=2$ are not compatible with existing
limits on \MD. 
The higher the $k$, the higher the limit on \Mth{} -- although the
highest $k$ may not be that chosen by nature.  
The $n=1$ case has its cross section modified to correspond to the RS1
model, and the limits are $(5.7\TeV,2.0\TeV)$ to $(5.0\TeV,4.0\TeV)$
at the 95\% CL.
Upper limits on $\sigma \times B \times A$ at the 95\% CL of about
0.2~fb$^{-1}$ are obtained.

ATLAS has only recently interpreted the dijet invariant mass spectrum in
terms of non-thermal black holes~\cite{Aad:2014aqa}.
Prior to this~\cite{Aad:2011aj,ATLAS:2012pu}, and in
Ref.~\cite{Aad:2014aqa}, the interpretation has been in terms of a
model that performs two-body Hawking evaporation at the Planck
scale~\cite{Dai:2007ki}. 
In such a model, the conservation principles followed in
Ref.~\cite{Calmet08b,Gingrich:2009hj} need not be obeyed
simultaneously. 
I consider this model as essentially the classical thermal black hole
model at the Planck scale.
Since this is the energy at which a classical model is anticipated not
to be valid, and the black hole must be treated as quantum mechanical
particle, I will not consider the model presented in
Ref.~\cite{Dai:2007ki} further. 

ATLAS interprets the non-observation of a broad enhancement in the dijet
invariant mass spectrum as a lower limit on the QBH mass of 5.66\TeV{}
at the 95\% CL for $\Mth = \MD$ and $n = 6$~\cite{Aad:2014aqa}.
Upper limits on the $\sigma \times A$ at the 95\% CL of about 0.2~fb
are obtained above a mass of about 5\TeV.
Recent results from ATLAS~\cite{atlas:run2dijet} using 80~pb$^{-1}$ of
data with the LHC running at 13\TeV{} centre of mass energy have
significantly improved the lower mass limit to 6.8\TeV.

\subsubsection{Searches in the photon and jets final state}

A similar search to dijets for QBH states can be performed in the photon
and jet invariant mass spectrum.
Since the photon is a electrically neutral vector particle, $QBH \to
\gamma +$parton decays are not allowed for all possible initial
parton--parton states. 
Possible initial states are $q+g$, $\bar{q}+g$, $q+\bar{q}$, and $g+g$.
The $u+g$ state dominates.
The $\gamma$+parton final state should appear in detectors as a
$\gamma +$jet invariant mass enhancement. 
To date, only one search by ATLAS has been performed~\cite{Aad:2013cva}. 
The signal model has been converted to a visible cross section and a
lower mass limit of 4.6\TeV{} at the 95\% CL has been obtained.
95\% CL upper limits on the visible cross section are obtained.
The visible cross section has a minimum of about 0.5~fb starting at a
mass of 3\TeV.

\subsubsection{Searches in the lepton and jets final state}

Strongly coupled gravity need not conserve global symmetries such as
baryon or lepton number. 
The $q+q$, $\bar{q}+\bar{q}$, $q + q^\prime$, and $\bar{q} +
\bar{q}^\prime$ states can produced $QBH\to \ell + q$, where $\ell$ is
a charged lepton, and $q$ is a quark or antiquark.
A gluon is not possible due to simultaneous conservation of total
angular momentum and electric charge.
This decay should appear in detectors as a lepton and jets.
Only ATLAS has performed a search~\cite{Aad:2013gma} by looking for an
enhancement in the $\ell+$jet invariant mass spectrum, where $\ell$ is
an electron or a muon. 
The signal model has been converted to a visible cross section and a
lower mass limit of 5.3\TeV{} at the 95\% CL obtained.
The 95\% CL upper limit on the $\sigma \times B$ is about 0.18~fb above
a mass of about 3.5\TeV.

\subsubsection{Searches in the dilepton final state}

Electrically neutral QBH states formed in $q+\bar{q}$ and $g+g$
collisions can decay to opposite signed dileptons.
Only one search by ATLAS~\cite{Aad:2014cka} has been performed by
looking in the dielectron and dimuon invariant mass spectra.
95\% CL upper limits on the $\sigma \times B$ are obtained.
The signal model has been converted to $\sigma \times B$ and a
combined lower mass limit at the 95\% CL of 3.65\TeV{} for an ADD model
and 2.24\TeV{} for an RS1 model have been obtained.
The $\sigma \times B$ at the 95\% CL is below about 0.5~fb above a mass
of about 2\TeV.

\subsubsection{QBH search summary}

Searches for QBH states in the dijet, $\gamma$+jets, $\ell+$jets, and
dilepton invariant mass spectra have been performed.
No enhancements have been observed and limits have been set on the
production cross section for these final states.
The cross section limits have been interpret in terms of lower limits on
the threshold mass.
All of the ATLAS searches have set lower mass limits using an
identical model and can be directly compared, as shown in the second
column of Table~\ref{tab1}. 
We see that the dijet limits are the highest in spite of the large QCD
background. 
The limits are restricted to the case $\Mth = \MD$ and $n=6$.

\begin{table}[htb]
\begin{center}
\begin{tabular}{|c|c|c|}\hline
Final state        & \multicolumn{2}{c|}{\Mth [TeV]}\\\cline{2-3}
                   & $\sqrt{s} = 8$ TeV{}& $\sqrt{s} = 13$ TeV\\\hline
jet + jet           & 5.7 & 8.5\\
$\ell$ + jet        & 5.3 & 7.5\\
$\gamma$ + jet      & 4.6 & 6.5\\
$\ell^+$ + $\ell^-$ & 3.6 & 5.0\\
\hline
\end{tabular}
\end{center}
\caption{95\% CL lower limits on the QBH threshold mass for different
final states in the ATLAS detector. The 8\TeV{} column corresponds to
published results, while the 13\TeV{} column are approximate predictions
for 3~fb$^{-1}$ of data.} 
\label{tab1}
\end{table}

\begin{figure}[htb]
\begin{center}
\includegraphics[width=0.45\textwidth]{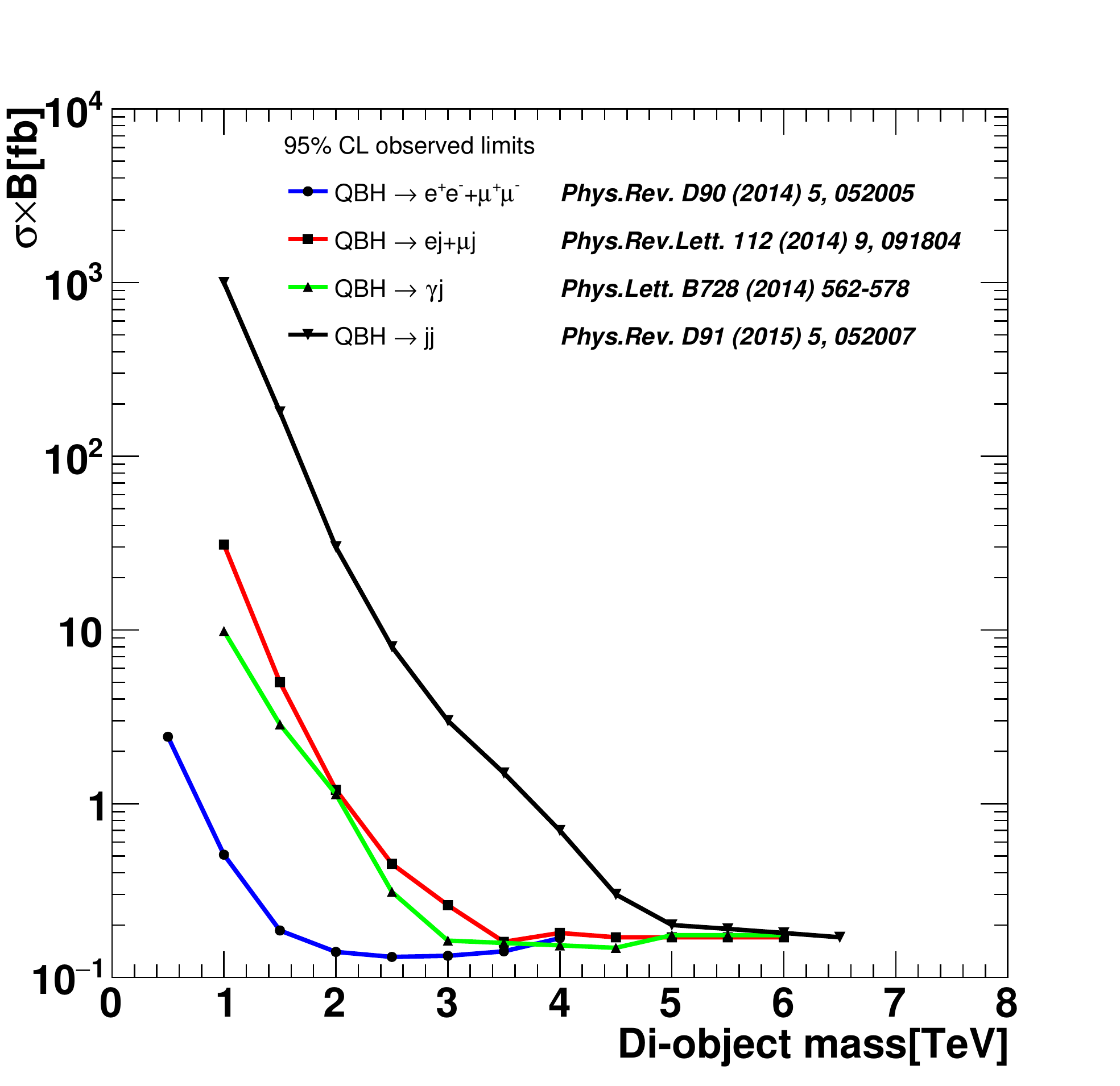}
\includegraphics[width=0.45\textwidth]{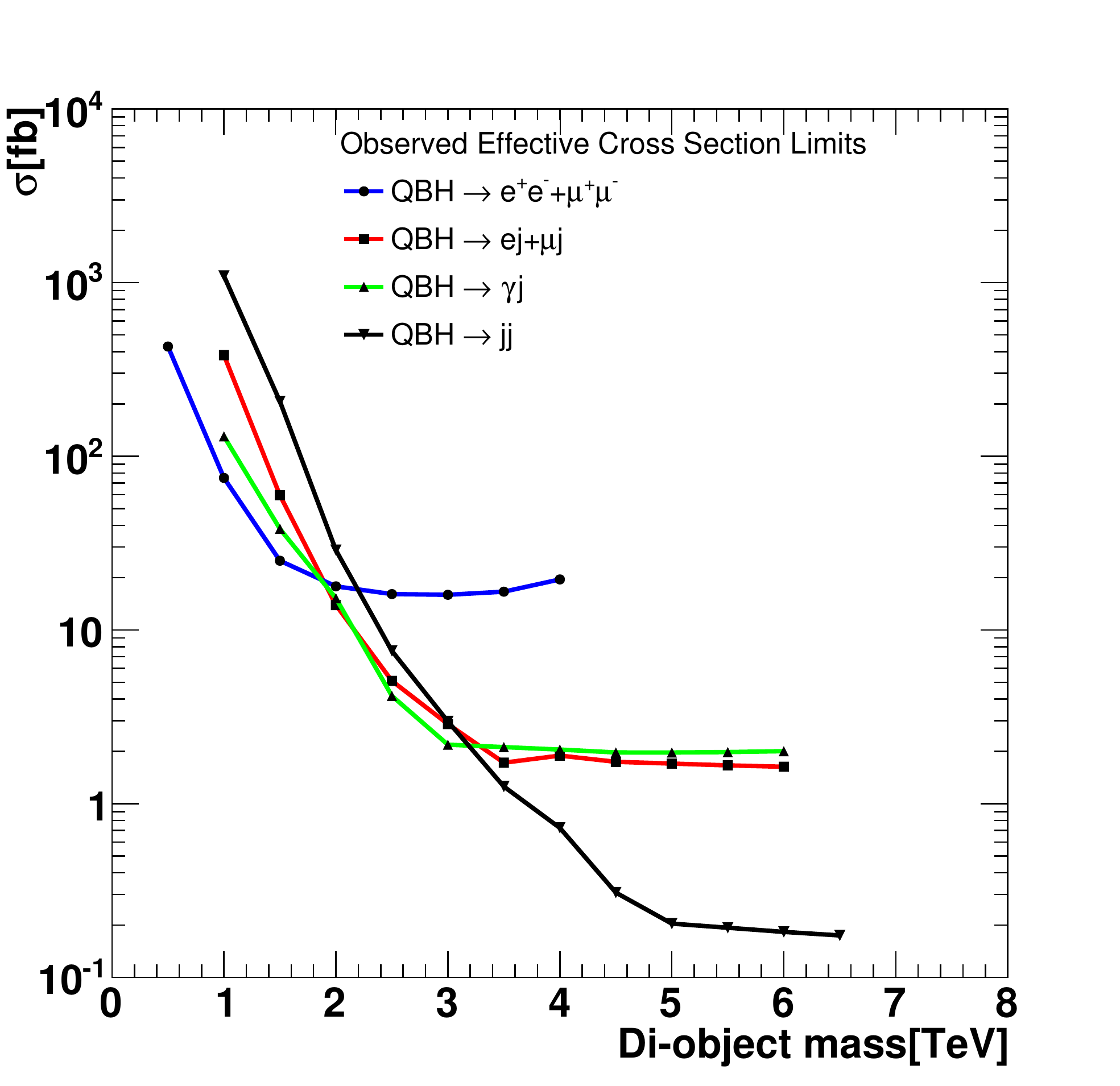}
\end{center}
\caption{ATLAS upper limits on (left) $\sigma \times B$ and (right)
$\sigma$ at the 95\% CL for QBH production in different final
states.} 
\label{fig3}
\end{figure}

The upper limits on $\sigma \times B$ can be compared as a function of
the two-object invariant mass. 
This is shown in Fig.~\ref{fig3}(left).
However, the branching fractions are different for each final state and
hence the limits can not be directly compared.
The same model used to set the limits can be used to estimate the
branching fractions.
Figure~\ref{fig3}(right) shows the combined limits on the production
cross section for QBH decaying to two objects.
Clearly, the dijet search provides the lowest cross section and
highest mass limits. 

\begin{figure}[p]
\begin{center}
\includegraphics[width=0.45\textwidth]{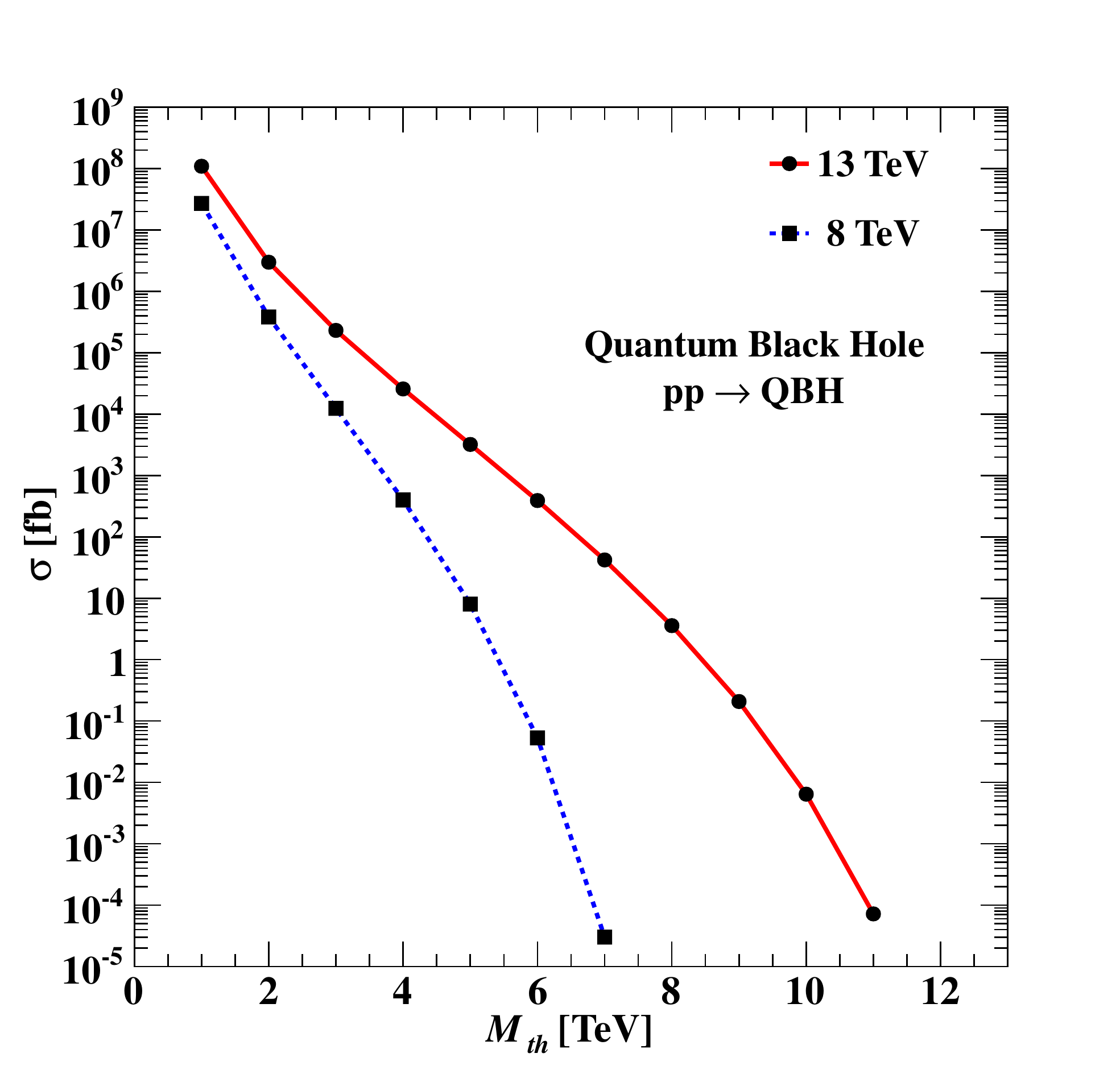}
\includegraphics[width=0.45\textwidth]{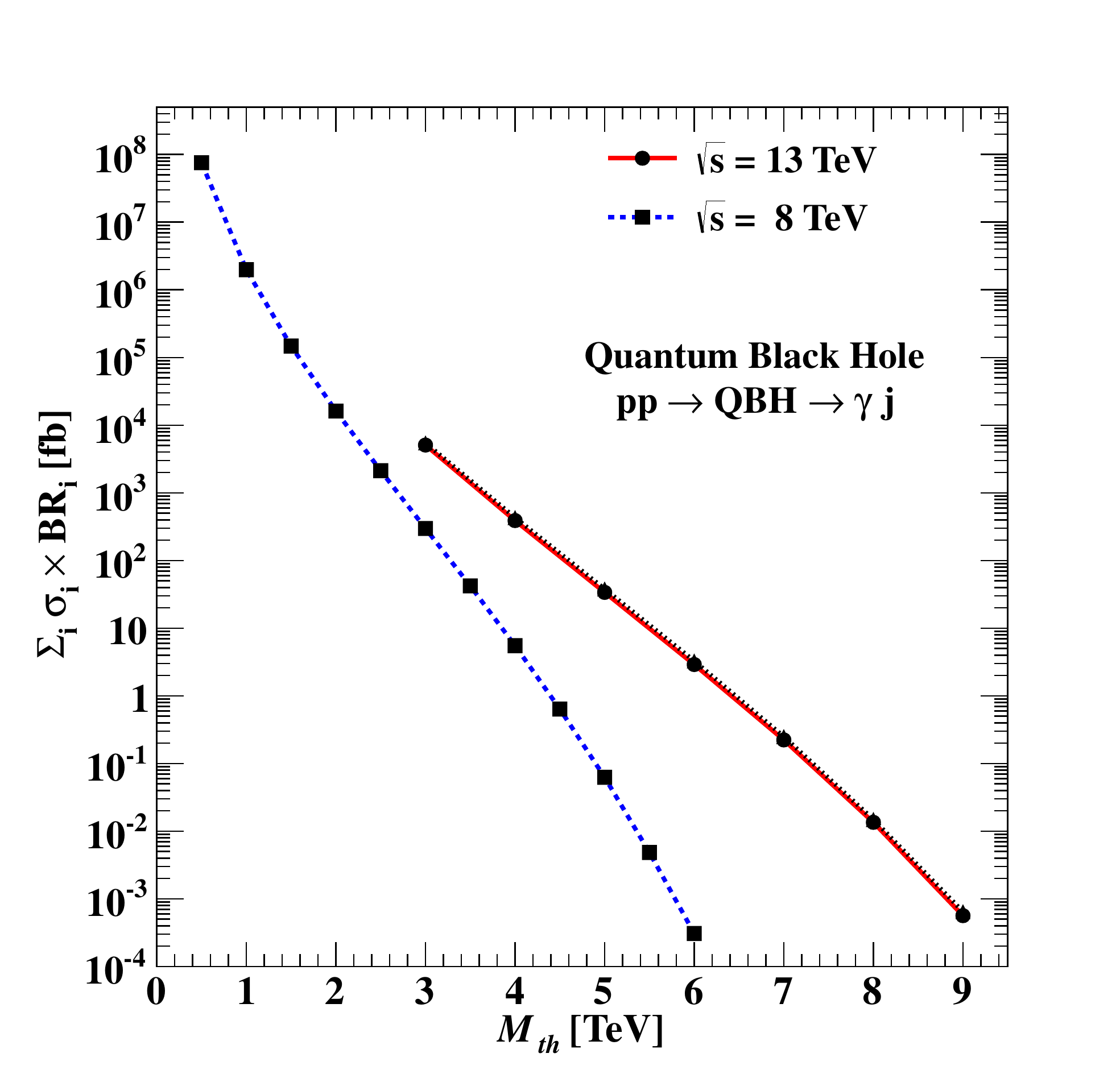}
\includegraphics[width=0.45\textwidth]{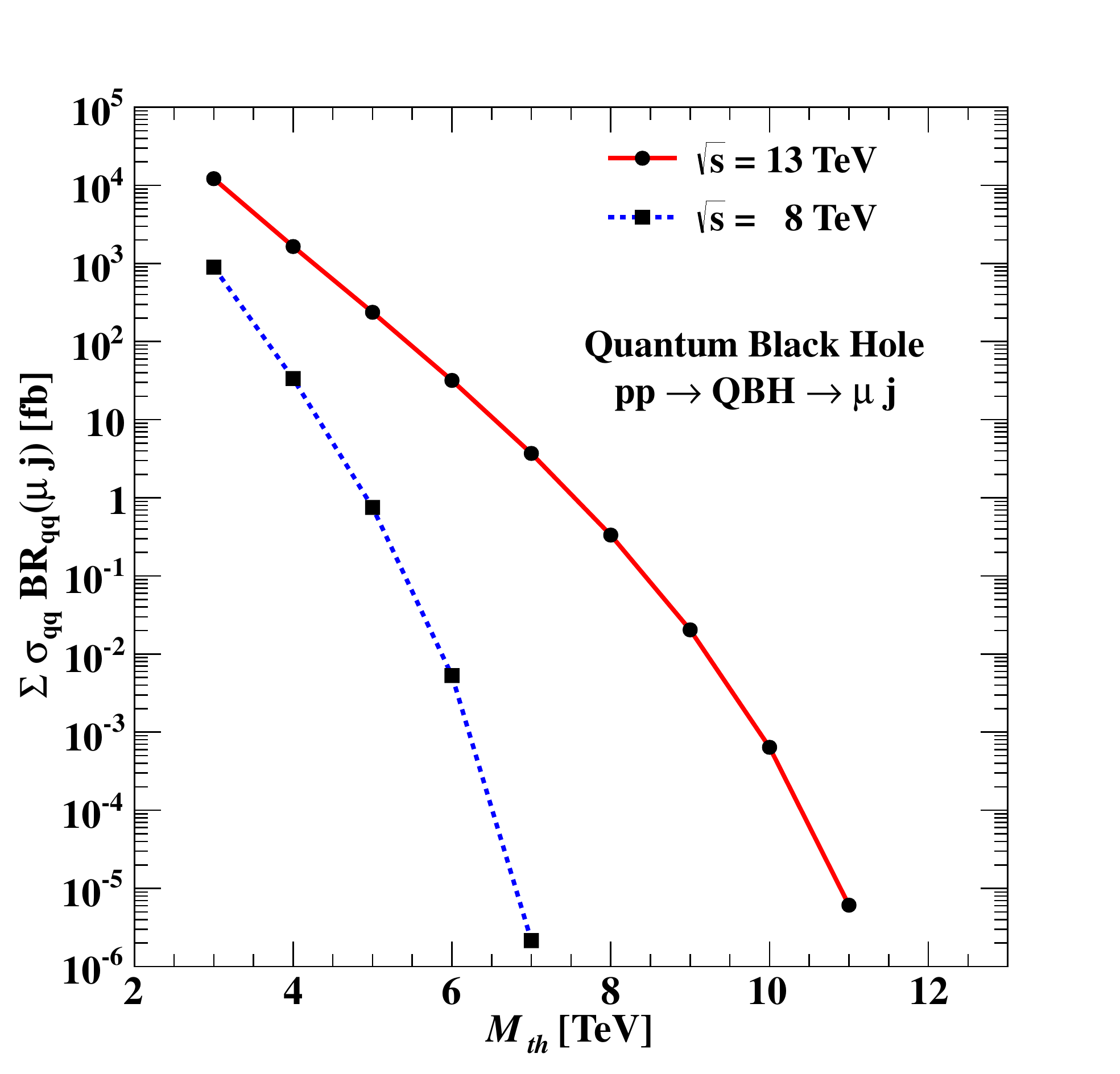}
\includegraphics[width=0.45\textwidth]{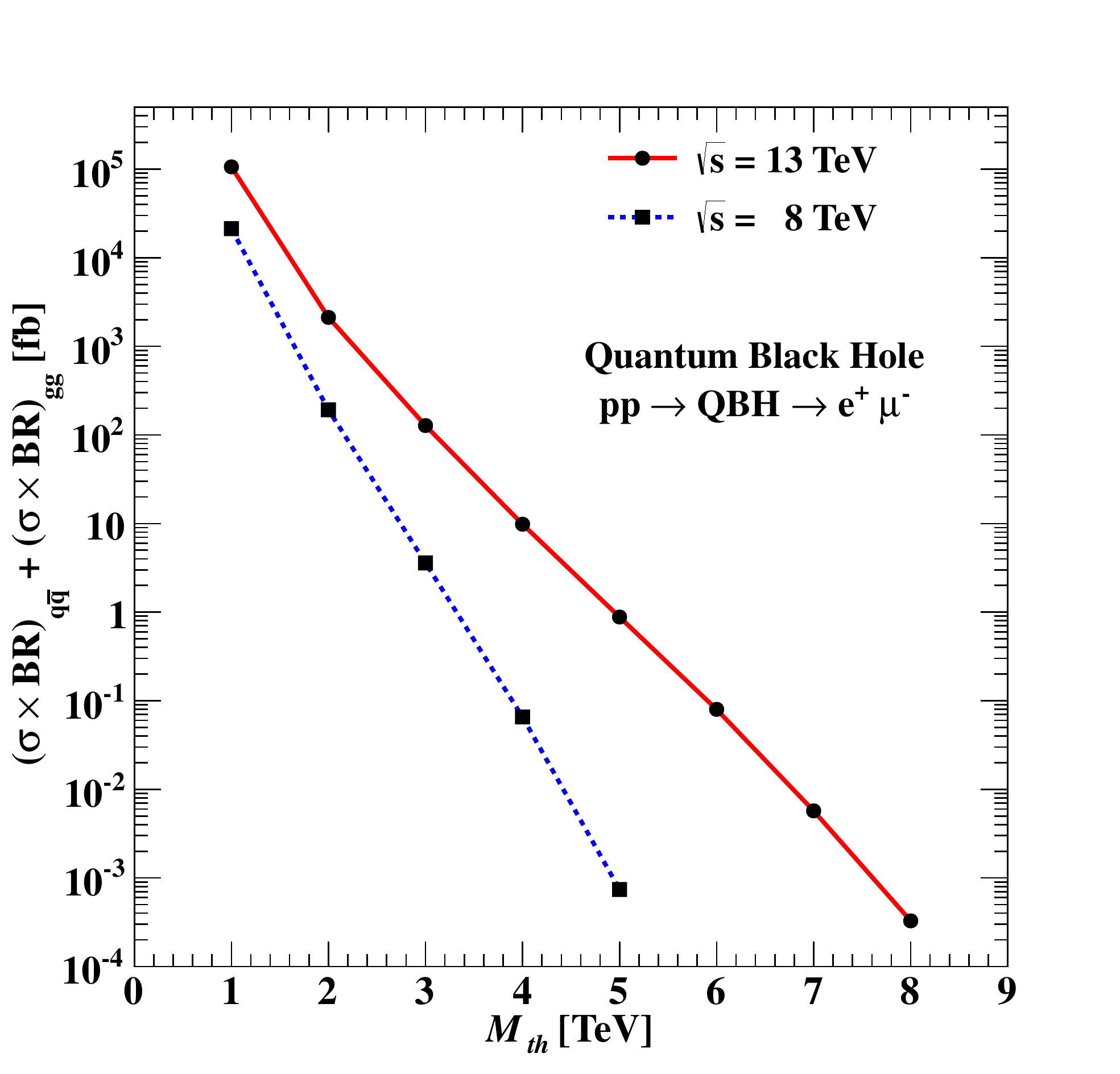}
\end{center}
\caption{Cross section time branching fraction for different QBH
channels at LHC proton--proton centre of mass energies of 8\TeV{} and
13\TeV: (top-left) inclusive, (top-right) $\gamma +$jet, (bottom-left)
$\ell+$jet, and (bottom-right) dilepton.} 
\label{fig4}
\end{figure}

Figure~\ref{fig4} shows the QBH production cross section time
branching fraction for proton--proton centre of mass energies of
8\TeV{} and 13\TeV.  
Significant mass sensitivity increases can be expected and are
estimated for 3~fb$^{-1}$ of data in the last column of
Table~\ref{tab1}. 

\section{What we think we know and alternatives}

Since low-scale gravity implies a new scale for physics, the search for
non-perturbative gravity is likely to be enhanced as the LHC energy
increases to the new scale.
Since the production cross sections are anticipated to be large above
the Planck scale, the usual view is that a search for non-perturbative
gravity is enabled by the highest energies -- not high luminosity.
If the LHC energy is near the new gravity scale, we might expect an
instant discovery at LHC turn-on at higher energies.
Of course this can be wrong and black holes can be produced at some low
rate at current energies, or decay to a different signature than that
searched for so far.
Two possibilities to reduce the cross section and make gravity
states difficult to detect, even if we are above the new physics
scale, are trap-surface calculations~\cite{Eardley02} and
split-fermion models~\cite{Arkani99p}. 
One of the only models that can predict new signatures, that I know 
of, is non-commutative geometry black hole models~\cite{Rizzo06b}. 
I will not discuss models that predict a stable remnant at the Planck
scale.

\begin{figure}[htb]
\begin{center}
\includegraphics[width=0.6\textwidth]{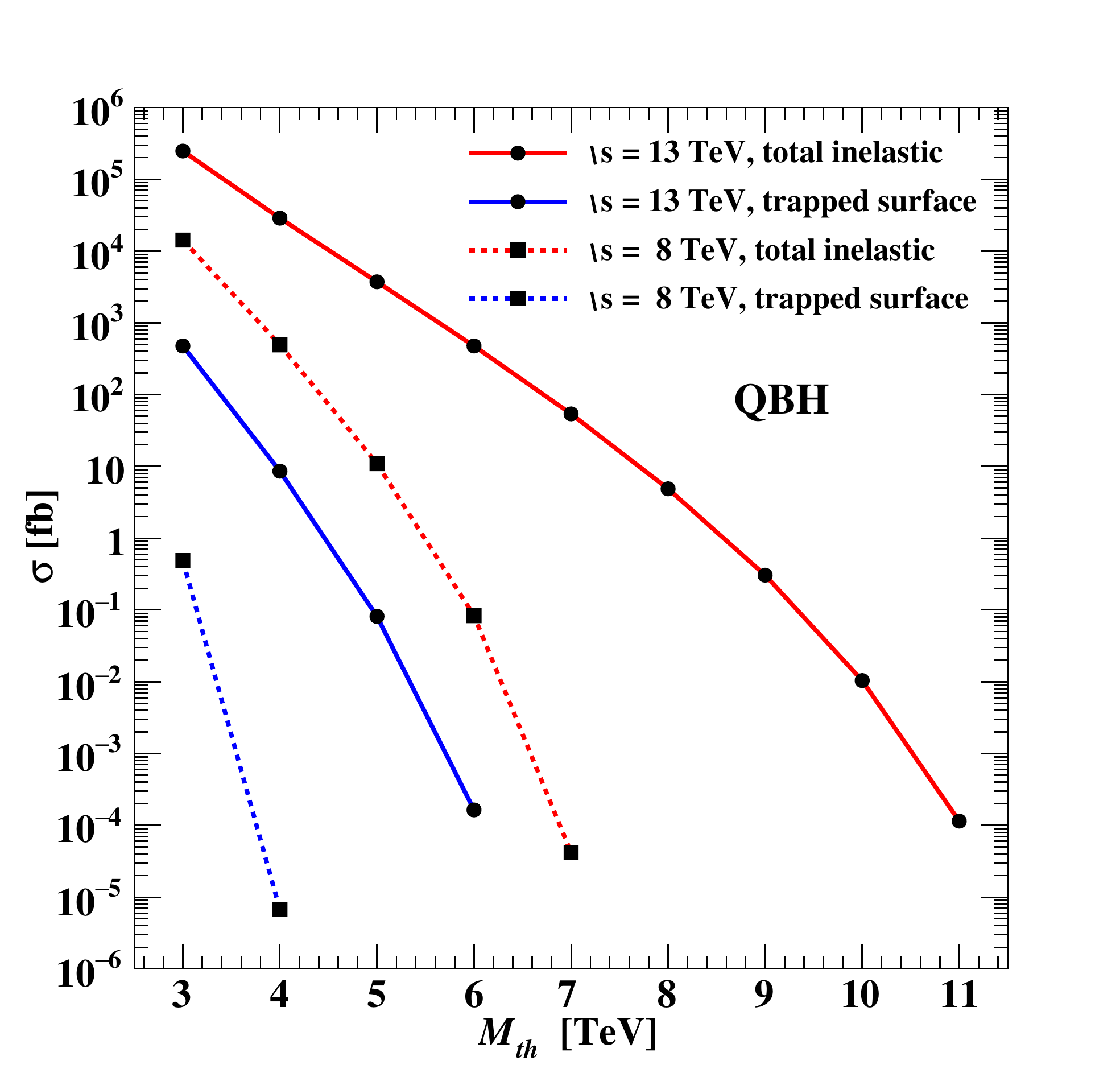}
\end{center}
\caption{Comparison of total inelastic cross section and trapped
surface cross section~\cite{Yoshino05a} applied to QBH production at
8\TeV{} and 13\TeV{} proton--proton centre of mass energy.}   
\label{fig5}
\end{figure}

Typically a total inelastic classical cross section form $\sigma = \pi
r_\mathrm{g}^2$, where $r_\mathrm{g}$ is the gravity radius is used
for the black hole parton--parton cross section. 
All the energy of the partons is assumed to go into producing the
black hole.  
But this is unlikely as various GR calculations predict only a
fraction of the energy in a particle--particle collision will be
trapped behind the horizon formed.
The excess energy ``appears'' as radiation.
I will refer to this initial-state radiation as radiation that can be
considered to occur before the black hole is formed, and balding
radiation as radiation that can be considered to occur after black
hole formation. 
In the former case, less energy is available for black hole formation
and the cross section is reduced. 
Neither of these radiation processes is consider as Hawking radiation.
Upper bounds on the amount of initial-state radiation for
higher-dimension black holes have been calculated~\cite{Yoshino05a}. 
Figure~\ref{fig5} shows the case of applying the trapped surface cross
section results from Ref.~\cite{Yoshino05a} to QBH production.

Split-fermion models are a mechanism for generating Yukawa hierarchies
by displacing the standard model fermion fields in a high-dimensional
space by localising them to different positions on a thick brane.
The overlap of the fermion wave functions give the couplings.
A set of spacings giving masses consistent with data has been
determined in a two-dimensional split-fermion
model~\cite{Branco:2000rb}. 
One can embed black holes and string balls in split-fermion models.
This can cause a reduction in cross section relative to the usual ADD
case~\cite{Abdolrahimi:2014bsa}.   
Split-fermion models have yet to be used to interpret results from the
LHC experiments.

ADD-type black holes have been embedded into a non-commutative
geometry~\cite{Rizzo06b}. 
It is hoped that some of the aspects of a theory of quantum gravity will
be reflected in the model by including the non-commutative geometry.
The matter distributions are smeared with a resolution given by the
non-commutative scale.
This introduces an extra parameter into the model: $\sqrt{\theta}$.
A benefit is that the temperature is well behaved during Hawking
evaporation all the way down to the Planck scale -- unlike the
semi-classical model in GR.
Or in other words, the canonical ensemble treatment of entropy is valid
for entire decay~\cite{Gingrich08a}. 
The gravitational radius has a non-zero minimum.
This results in a remnant.
But unlike most remnant models, the remnant mass is different from the
Planck scale. 
The model exists and gives rather different phenomenological
signatures than the usual models~\cite{Gingrich:2010ed}.  
The main experimental differences include a large missing energy in
events and a soft \sumpT{} spectrum of particles.

\section{Discussion}

For the model-independent analyses, upper limits on the fiducial cross
section as a function of inclusive \sumpT{} are determined. 
There seems no good method for removing the model dependence and making
the results generic. 
In an ideal world, it would be beneficial to have the model dependence
removed from the model-independent limits.
Especially for low-scale gravity in which the models are rather
speculative.

Model-dependent limits are set in the two-dimensional parameter space
of \Mth{} and \MD.
The other parameters are fixed and the result is called a model.
The choice of which parameters to leave fixed and their values are
somewhat arbitrary.
For example, $\gs = 0.4$ is held constant when setting limits on string
balls.
The choice of \gs{} can case the search to be sensitive in the
stringy, unitarity, or black hole regime of cross section depending
on its value. 
Thus the limits may depend significantly on the choice of \gs.
Lower mass limits are set on \Mth{} as a function of the other
parameters in the model.
I remind the reader that \Mth{} is not a physical parameter but a
cut-off to parameterise the validity of the model.
\Mth{} can vary considerably over the range of \MD{}, and presumably over
the possible range of other parameters that are held fixed at arbitrary
values.
At best, model-dependent limits allow a comparison of the sensitivity
to different models relative to each other in a given analysis strategy,
and a comparison of the sensitivity to the same model between analysis
strategies. 
I see no physics reason for producing model-dependent limits.
 
In most cases, searches for thermal states are performed in the \sumpT{}
variable. 
This variable is not directly related to the analytical form of the cross
section. 
For the model-independent limits, it may be possible to obtain higher
mass limits by using invariant mass rather than \sumpT.
It may even be possible with some additional work to perform the
search in \sumpT, but set the limits in mass.
Cross section limits in mass would allow a direct use of the theory to
determine the upper mass reach, and possibly limits on \MD.
This is already done in the non-thermal limits which are preformed in
mass. 

The analyses in ATLAS dealing with thermal black hole searches do not
consider missing transverse momentum.
This seems puzzling since neutrinos give rise to missing momentum and
should be produced with equal, or one-half, the probability of charged
leptons. 
If the production of charged leptons is significant enough to design a
search based on them, the effect of missing energy from neutrinos should
be of some significance in the analysis if not accounted for. 
In addition, in a model of low-scale quantum gravity it is hard to
imagine that gravitons are not produced.
Gravitons should give rise to missing energy, and their production may
well be significant~\cite{Gingrich07a}.

\section{Summary}

To date, the experiments at the LHC have published a total of 16 papers
which search for non-perturbative gravity states.
Based on lower limits on the Planck scale and an acceptable validity of
the GR black hole model, these states have been ruled out at LHC
energies of $13-14\TeV$. 
By the same reasoning, string ball states are also likely to be ruled out,
but there may be a small window of validity above 13\TeV.

Searches for non-thermal black holes allow a direct model-dependent
limit of the mass.
The dijet channel is very powerful and sets the most stringent limits
on the low-scale gravity scale. 
It would be beneficial to use the dijet signature to interpret models
that predict significantly lower cross sections then the usual models --
like RS1 and trapped surface cross sections.

Low-scale gravity studies are expected to benefit more from an increase
in LHC energy than luminosity. 
This expectation is based on the nominal models.
Quantum gravity effects, or other unaccounted for effects, may cause
cross sections to be lower then expected.

Prior to the turn-on of the LHC a large number of papers discussing
low-scale gravity were written.
The overwhelmingly majority used 1\TeV{} as the gravity scale.
In spite of some short comings of the non-thermal state searches,
combined with the limits on \MD{} from mono-jets, it is advisable to
increase the gravity scale to at least 3\TeV{} in any future
phenomenological studies.
The predictions in proton--proton collisions will be less dramatic
than those early papers of about 17 years ago.

\section*{Acknowledgments}

I would like to acknowledge useful discussions with the following
colleagues:
James Dassoulas,
Jack Edwards, 
Kuhan Wang, and
Zihui Wang.

This work was supported in part by the Natural Sciences and Engineering
Research Council of Canada.

\bibliographystyle{atlasBibStyleWithTitle}
\bibliography{gingrich}

\end{document}